\documentclass[10pt,letterpaper]{article} 
\usepackage{titlesec,float,makeidx,multirow,multicol,graphicx,hyperref,amsmath,amssymb,fancyhdr,changepage,etoolbox,times,enumitem,textcomp,indentfirst,subcaption,caption,setspace}

\usepackage[dvipsnames,svgnames,table]{xcolor}
\usepackage{txfonts}
\usepackage{mathdots}
\usepackage[paperwidth=8.5in,paperheight=11in,top=1in,right=0.5in,bottom=1in,left=0.5in]{geometry}
\usepackage[utf8]{inputenc}
\usepackage[sort&compress,numbers]{natbib} 
\fancypagestyle{firstpage}{
\fancyhead[L,C,R]{}
\fancyfoot[R]{ }
\fancyfoot[C]{\thepage}
\fancyfoot[L]{}

}

\RequirePackage{fix-cm}
\newcommand{\size}[2]{{\fontsize{#1}{0}\selectfont#2}}

\titleformat{\section}
  {\normalfont\fontfamily{phv}\bfseries}{\thesection}{10pt}{\MakeUppercase}
\renewcommand*{\thesection}{\fontfamily{phv}\selectfont\textbf{\arabic{section}.}}
\titlespacing{\section}{0pt}{10pt}{0pt}
\titleformat{\subsection}
  {\normalfont\fontfamily{phv}\bfseries}{\thesubsection}{3pt}{}
\renewcommand*{\thesubsection}{\fontfamily{phv}\selectfont\textbf{\arabic{section}.\arabic{subsection}}}
\titlespacing{\subsection}{0pt}{10pt}{0pt}

\DeclareCaptionFormat{upper}{#1#2\uppercase{#3}\par}
\newcommand{\startsquarepar}{%
    \par\begingroup \parfillskip 0pt \relax}
\newcommand{\stopsquarepar}{%
    \par\endgroup}

\hypersetup{
	colorlinks=true,
	linkcolor=black,
	filecolor=blue,      
	urlcolor=blue,
	citecolor=black,
}
\usepackage{nomencl}
\usepackage{bm}
\usepackage{xstring}
\usepackage{xpatch}
\patchcmd{\thenomenclature}
  {\leftmargin\labelwidth}
  {\leftmargin\labelwidth\itemindent 0.25in }
  {}{}

\makenomenclature

\renewcommand\nomgroup[1]{%
\item[\bfseries
\ifstrequal{#1}{A}{Fluid and flow parameters}{%
\ifstrequal{#1}{B}{Geometric parameters}{%
\ifstrequal{#1}{C}{Subscripts}{}}}%
]}
\begin{document}
\newgeometry{top=0.5in,right=0.5in,bottom=1in,left=0.5in}
\begin{flushright}
\fontfamily{phv}\selectfont{
\textbf{Conference Paper}\\
\textbf{2022}\\
\textbf{ }\\
\textbf{ }\\[45pt]
\textbf{\size{18}{ }}\\[38pt]
}
\end{flushright}
\begin{center}
    \fontfamily{phv}\selectfont{\size{11}{\textbf{Shock System Dynamics of a Morphing Bump Over a Flat Plate\\[20pt]}}}
\end{center}

    \begin{flushright}
        \fontfamily{phv}\selectfont{\textbf{Ahmed A. Hamada$^{1,2}$, Lubna Margha$^{1,2}$, Mohamed M. AbdelRahman$^{2}$, Amr Guaily$^{3,2}$} \newline
        $^1$Department of Ocean Engineering, Texas A$\&$M University, College Station, TX, 77843, USA \newline 
        $^2$Department of Aeronautical and Aerospace Engineering, Cairo University, Giza, 12613, Egypt \newline
        $^3$Smart Engineering Systems Research Center (SESC), Nile University, Giza, 12588, Egypt \newline}
    \end{flushright}
\begin{multicols*}{2}
\section*{Abstract}
\textit{The shock wave boundary layer interaction (SW-BLI) phenomenon over transonic and supersonic airfoils captured the attention of aerospace engineers, due to its disastrous effect on the aerodynamic performance of these vehicles. Thus, the scientific community numerically and experimentally investigated several active and passive flow control elements to reduce the effect of the phenomenon, such as vortex generator, cavity, and bump. They focused on designing and optimizing the shape and location of the bump control element. However, the transit movement of the bump from the state of a clean airfoil to the state of an airfoil with a bump needs more investigation, especially the dynamics of the shock system. Thus, it is preferred to start with simple geometry, such as a flat plate, to fully understand the flow behavior with a morphing bump. In this paper, the shock dynamics due to the movement of a bump over a flat plate flying at supersonic speed are numerically investigated. The bump is located at the impingement position of the shock wave and is moved at different speeds. This study determines the suitable speed that achieves the minimum entropy change, which is the representation parameter of the transition period. The two-dimensional unsteady Navier-Stokes equations are solved using OpenFOAM to simulate the flow field variables, while the motion of the bump is tracked using the Arbitrary Lagrangian-Eulerian (ALE) technique. The results show that a spatial lag on the shock system from the steady-state solution occurs due to the movement of the bump. Further, the spatial lag increases with the increase in the bump's speed. This causes a high increase in the flow parameters and consequently the total entropy changes on the bump surface. Generally, it is common to move the bump over the longest possible time to approximate a quasi-steady flow during the motion. However, this causes a deviation in the flow parameters between the final time of transition and the steady-state case of bump existence. Thus, it is concluded that the optimal non-dimensional time for a morphing bump in a supersonic flow of Mach number of 2.9 is 2, which is different than the longest time of 10.}

Keywords: Moving bump; Supersonic flow; Flat plate; Active flow control. 
\mbox{}
\nomenclature[A]{$\tau$}{Non-dimensional time}
\nomenclature[A]{$M$}{Mach number}
\nomenclature[A]{$Re$}{Reynolds number}
\nomenclature[A]{$e$}{Specific energy per unit mass}
\nomenclature[A]{$s$}{Entropy}
\nomenclature[A]{$t^{\ast}_f$}{Motion period}
\nomenclature[A]{$\mu$}{Dynamic viscosity}
\nomenclature[A]{$\rho$}{Density}
\nomenclature[A]{$p$}{Pressure}
\nomenclature[A]{$T$}{Temperature}
\nomenclature[A]{$c_v$}{Specific heat at constant volume}
\nomenclature[A]{\textit{\textbf{V}}}{Velocity vector}
\nomenclature[B]{$c$}{Chord}
\nomenclature[B]{$h$}{Height}
\nomenclature[B]{$l$}{Length}
\nomenclature[B]{$\alpha$}{Height to length ratio}
\nomenclature[C]{$b$}{Bump}
\nomenclature[C]{$i$}{Inlet}
\nomenclature[C]{$sl$}{Slip wall}
\nomenclature[C]{$w$}{Wedge}
\nomenclature[C]{$\infty$}{Free-stream}
\printnomenclature[0.68in]
\section{Introduction}
\startsquarepar \indent The Shock wave boundary layer interaction (SWBLI) over a high-speed wing disastrously affects its aerodynamics performance and structural lifetime. The phenomenon of SWBLI was experimentally observed for the first time by Ferri \cite{ferri1939experimental} in 1939. Green \cite{green1970interactions} summarized the occurrence of the phenomenon in four conditions: externally as in transonic airfoils, and near control flaps, or internally as in high-speed inlets (scram-jets), and nozzles at off-design. Based on strength of the shock wave, the \stopsquarepar
\end{multicols*}
\restoregeometry
\clearpage
\begin{multicols*}{2} 
\noindent phenomenon varies the height of the boundary layer, the surface drag of the body, and/or the heat transfer. The SWBLI adversely affects the structure and geometry of the flying vehicles \cite{dolling2001fifty}. The flow experiences a high instantaneous increase in pressure and thermal transfer due to the existence of SWBLI, which leads to a reduction in the fatigue life of the structure. Consequently, that imposes strong constrictions on choosing the material of the structure, resulting in an expensive and heavy design. Moreover, this phenomenon happens in transonic airfoils after a critical Mach number, causing the sudden increase in total drag, which is called “drag divergence”. 

There are several ways to eliminate or reduce the SWBLI effect which is discussed in detail by Dennis Bushnell \cite{bushnell2004shock}. The focus in this paper will be on the bump approach. In 1922, Ashill et al. \cite{ashill199292} were the first researchers who used a two-dimensional bump at the upper surface of airfoils to affect the strength of incurring shock waves. Milholen et al. \cite{milholen2005application} and Patzold et al. \cite{patzold2006numerical} proved that using bumped airfoils increases the lift, reduces the drag, and postpones the buffeting, which enhances the aerodynamic performance. Fulker \cite{fulker1999euroshock}, who is a DASA-Airbus researcher, showed that applying a bump in A-340’s hybrid laminar wing will save fuel by $2.11\%$ at $M_{\infty}=0.84$. Several investigations are conducted to conclude with the optimal shape and location of the bump \cite{sommerer2000numerical,wong2007parallel,tian2014multi,mazaheri2015optimization}. Further, Eastwood and Jarrett \cite{eastwood2012toward} investigated the three-dimensional bump control technique. Yun Tian et al. \cite{tian2011shock} showed that airfoil geometry and free stream condition determine the bump's optimal geometric parameters which specify its shape and location over the airfoil. In addition, varying these geometric parameters highly affects the aerodynamic performance.


By the beginning of the 21st century, the morphing wing concept become industrial applicable. Stanewsky et al. discussed the design aspects of the morphing concept from the view of the civil aircraft industry \cite{monner2000design}. From that time, many scholars started to investigate the flow behavior over the morphing wing at different velocity regimes \cite{olivett2021flow,elbadry2021active,botez2018numerical,popov2010real}. Further, Bruce and Colliss \cite{bruce2015review} conducted a review study for the shock control bumps, showing that the morphing bump is a future ideal solution to limit the SWBLI effect on wings at transonic flows. 

\begin{figure}[H]
	\centering
	\includegraphics[width=\linewidth]{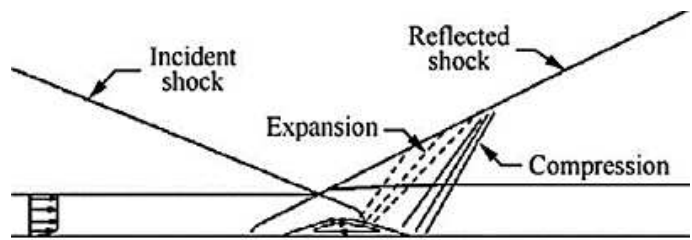}
	\caption[A schematic of 2D SW-BLI separation flow.]{A schematic of 2D SW-BLI separation flow [21].}
	\label{fig:Fig0}
\end{figure}

To simplify the phenomenon, consider a strong two-dimensional shock wave that impinges with a boundary layer of a flat plate, see Figure \ref{fig:Fig0}. Downstream the impinging region, separation, and relaxation zones are generated. The flow changes from a non-equilibrium state in the separation zone to an equilibrium state in the relaxation zone. Through these zones, the flow experiences a quick increase in pressure and heat transfer, detaching, and reattaching the boundary layer. Moreover, the turbulent kinetic energy increases due to the generation of the strong shear and the adverse pressure gradient in and near the separation zone, respectively. During the phenomenon, the flow contains different types of waves generated from the SW-BL.I, such as; a reflected shock wave, expansion waves, and compression waves \cite{ma2014study}.

The static condition for the bump control element is fully investigated. Particularly, they concentrated on designing and optimizing the bump control element's shape and position. However, additional detailed investigations are needed for the transition of the morphing bump from a clean airfoil to an airfoil with a bump, focusing on the dynamics of the shock system. The aim of this research is to numerically study the transient effect of the morphing pump over a simple geometry (flat plate) on the unsteady flow features of the shock dynamics. We started to study the transient phenomenon over a simple geometry first to fully understand the transition effects. The location of the bump is chosen to be at the impingement location of the incident shock wave. Then, starting the unsteady motion of the bump at different speeds and a constant velocity profile. This enabled us to determine the suitable morphing bump's speed that would minimize the entropy change.

\section{Physical Model}
\subsection{Model Description}
\begin{figure}[H]
	\centering
    \includegraphics[width=\linewidth]{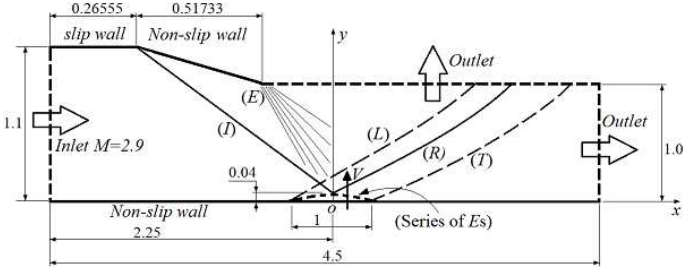}
    \caption{ Schematics of shock system over the morphing bump.}\label{fig:Fig1}
\end{figure}

\begin{figure}[H]
	\centering
    \includegraphics[width=\linewidth]{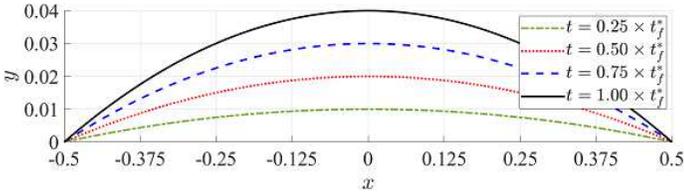}
    \caption{ Shape of the morphing bump with time.}\label{fig:bump_time}
\end{figure}

The flow configuration of the morphing bump in a supersonic flow is shown in Figure \ref{fig:Fig1}. Further, the variation of the morphing bump shape with time is shown in Figure \ref{fig:bump_time}. The height of the computational inflow boundary is $1.1m$, and the height of the outflow at the right is $1m$. The difference in height between the two edges represents the height of the wedge, which is the shock-source surface. The wedge is far from the inlet with a length of $0.2656m$ and the wedge stream-wise length is $0.5173m$. The morphing bump is placed at the center of the bottom surface, representing the origin of the domain. The origin is $2.25m$ far away from the inlet. The final shape of the morphing bump is a parabolic arc of height-to-length ratio, $\alpha= 4\%$, and represented by Equation (\ref{eq:mp}).
\begin{equation} \label{eq:mp}
y=\sqrt{\left(\frac{\alpha^2+0.25}{2\alpha}\right)^2-x^2}+\frac{\alpha^2-0.25}{2\alpha}
\end{equation}

\indent When the supersonic flow with a free-stream Mach number $M_\infty= 2.9$ and a free-stream Reynolds number $Re_{\infty}=6.6\times10^{7}$ impinges the wedge, an incident shock wave (\textit{I}) is generated hitting the bottom surface. Then, a reflected shock wave (\textit{R}) appears. Further, a series of expansion fans generates from the wedge trailing edge (\textit{E}). The expansion fan interacts with the shock system causing a curvature to them. The remaining top surface after the wedge is considered an outlet to enable the flow to go out freely. The morphing bump rises to its final shape with a constant velocity profile and different speeds (different motion periods of $0.1$, $1.0$, $5.0$, $7.5$, $10.0$). The initial condition of the problem is shown in Figure \ref{fig:Fig1} (a) with the solid lines, while the dashed lines are for the final condition. The values of the problem's parameters are shown in Table \ref{table:1}. The appearance of the morphing bump causes a generation of two oblique shocks at its leading edge (\textit{L}) and at its trailing edge (\textit{T}). Further, a series of expansion fans appear over the morphing bump due to its geometric curvature (Series of \textit{E}s).
\begin{table}[H]
\centering
\caption{System properties and parameters.}
\begin{tabular}{c c} 
 \hline \rule{0mm}{2.5ex}
Mach number, $M_{\infty}$         & $2.9$               \\[0.1ex]
 \hline
Reynolds number, $Re_{\infty}$         & $6.6\times10^7$             \\[0.25ex]
 Inlet length, $l_i$         & $1.1 m$             \\[0.1ex]
 Wedge chord, $c_w$              & $0.527 m$      \\[0.1ex]
 Wedge angle, $\alpha_w$              & $11^{\circ}$      \\[0.1ex]
 Slip wall length, $l_{sl}$              & $0.26555 m$      \\[0.1ex]
 Bump final length, $l_b$ & $0.5 m$      \\[0.15ex]
 Bump final height, $h_b$ & $0.01 m$\\[0.15ex]
 Motion periods, $t^\ast_f$        & $0.1$, $1$, $1.8$, $2$, $3$, $5$, $7.5$, $10$\\[0.5ex]
\hline
\end{tabular}
\label{table:1}
\end{table}
\subsection{Governing Equations}
Two-dimensional unsteady compressible Navier-Stokes equations are used to model the supersonic flow, and are expressed as:
\begin{subequations}\label{eq:GEq}
\begin{align} 
\frac{\partial \rho}{\partial t}&+\nabla \cdot \left(\rho \textbf{\textit{V}}\right) =0\\
\rho \left(\frac{\partial \textbf{\textit{V}}}{\partial t}+\left(\textbf{\textit{V}}\cdot\nabla\right)\textbf{\textit{V}}\right)&\nonumber\\=-\nabla p+\nabla \cdot &\left(\mu\left(\left(\nabla\textbf{\textit{V}}\right)+\left(\nabla\textbf{\textit{V}}\right)^T\right)\right) -\frac{2}{3} \nabla\left(\mu\nabla \cdot \textbf{\textit{V}}\right)\\
\rho \left(\frac{\partial e}{\partial t}+\left(\textbf{\textit{V}}\cdot\nabla\right)e\right)&+p\left(\nabla\cdot\textbf{\textit{V}}\right)\nonumber\\ = -\frac{2}{3}\mu&\left(\nabla\cdot\textbf{\textit{V}}\right)^2+\mu\left(\left(\nabla\textbf{\textit{V}}\right)+\left(\nabla\textbf{\textit{V}}\right)^T\right): \left(\nabla\textbf{\textit{V}}\right)
\end{align}
\end{subequations}
where $\rho$ is the density, $V$ is the velocity vector, $p$ is the pressure, $e$ is the specific energy per unit mass, $\mu$ is the fluid dynamic viscosity.

\section{Computational Model}
\subsection{Numerical Implementation}
\textit{rhoCentralDyMFoam} is a toolbox flow solver in OpenFOAM\textsuperscript{\textregistered}$-$v2006. It solves transient, compressible flow problems using a density-based approach, supporting dynamic mesh applications. The semi-discrete and upwind-central non-staggered schemes of Kurganov and Tadmor are combined in the solver \cite{kurganov2000new,kurganov2001semidiscrete}, which uses an operator-splitting method \cite{greenshields2010implementation}. The van Leer limiter is implemented in the solver to balance between shock capture, oscillations-free fields, and computational cost during the solution \cite{marcantoni2012high}.

The letters "\textit{DyM}" in the name of the \textit{rhoCentralDyMFoam} solver indicates that the dynamic mesh simulation is applicable with a dynamic FV mesh technique. In this study, the motion of the bump will be actively controlled. Thus, the \textit{coded} motion solver is used in this study to arbitrarily move the computational mesh inside the domains, optimizing the cells' shape. Hence, the dynamic FV mesh follows the Arbitrary Lagrangian-Eulerian technique, which is presented by Hirt et al. \cite{hirt1974arbitrary}.
\subsection{Computational Domain}
\begin{figure}[H]
	\centering
    \includegraphics[width=0.99\linewidth]{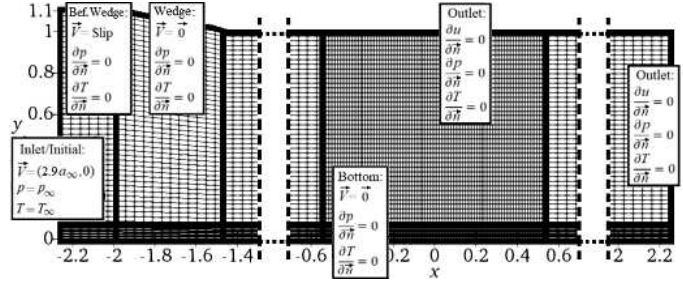}
    \caption{  Schematics of the computational domain for mesh 1.}\label{fig:Fig2}
\end{figure}

A body-fitted computational domain is constructed without the existence of the bump. The domain is clustered in the direction of the $y$-axis along the bottom surface and also clustered in the $x$-direction at the area of the morphing bump. The domain was divided into 10 blocks where the cells are nearly orthogonal and quadrilateral. The grid will be updated each time step during the motion of the bump, preserving the body-fitted mesh by applying the Arbitrary Lagrangian-Eulerian (ALE). Four types of boundaries define the borders of the domain: inlet, outlet (right and top after the wedge), non-slip walls (bottom and wedge), and slip wall (top before wedge). Figure \ref{fig:Fig2} shows the boundary and initial conditions over the Schematic of the grid.

\begin{figure}[H]
    \centering
    \includegraphics[width=0.99\linewidth]{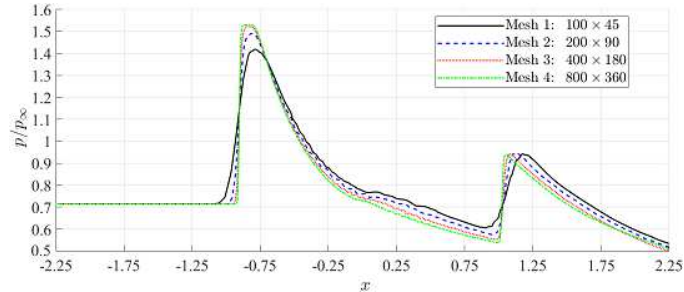}
    \caption{Mesh independent study for a supersonic flow over a clean flat plate, by comparing the pressure distribution at the height, $y=0.5m$.}\label{fig:Fig3}
\end{figure}

A mesh-independent study is essential to ensure the result's accuracy. The study was implemented over the case of without bump, starting with mesh 1 of size $100\times 45$ cells. Then, high-density meshes (2, 3, and 4) were generated by doubling the number of cells in $x$ and $y$ directions. The time step was controlled with the Courant–Friedrichs–Lewy (CFL) number of $0.2$. Figure. \ref{fig:Fig3} shows the variation of the pressure distribution at height of $y = 0.5m$ for different mesh sizes. The results with mesh 3 show a converged solution that does not depend on the mesh quality. This is because the difference between mesh 3 and mesh 4 is negligible. Thus, Mesh 3 of size $400\times 180$ was chosen to implement all the simulations to minimize computational time. The minimum element size in mesh 3 is $2.8mm \times 1.2mm$, and the time step to ensure the CFL number of $0.2$ is $111.1 \mu s$.
\section{Results and Discussion}
\subsection{Shock Structure over a Stationary Bump}
\begin{figure}[H]
	\centering
	\begin{subfigure}{\linewidth}
		\centering
		\includegraphics[width=\linewidth]{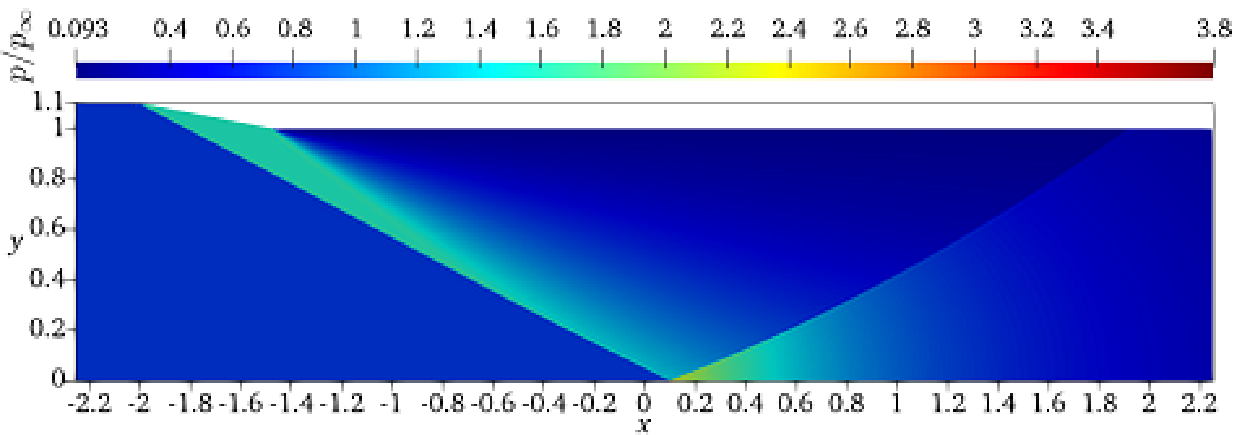}
		\caption{ Without a bump.}\label{fig:Fig5}
	\end{subfigure}
	\begin{subfigure}{\linewidth}
		\centering
		\includegraphics[width=\linewidth]{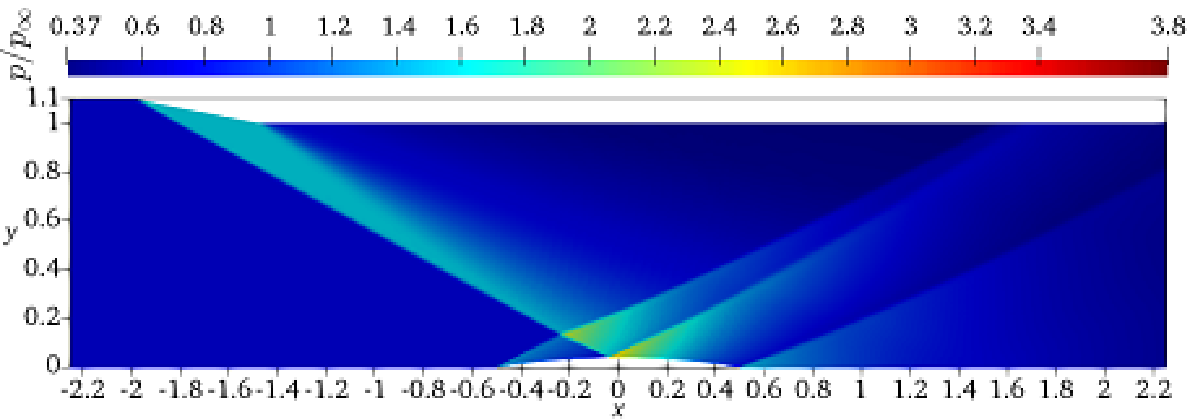}
		\caption{ With a stationary bump.}\label{fig:Fig6}
	\end{subfigure}
	\caption{ Pressure contours of supersonic flow over a flat plate with and without a bump at dimensionless time of 10.}\label{fig:Fig5_6}
\end{figure}

\begin{figure}[H]
    \centering
    \includegraphics[width=\linewidth]{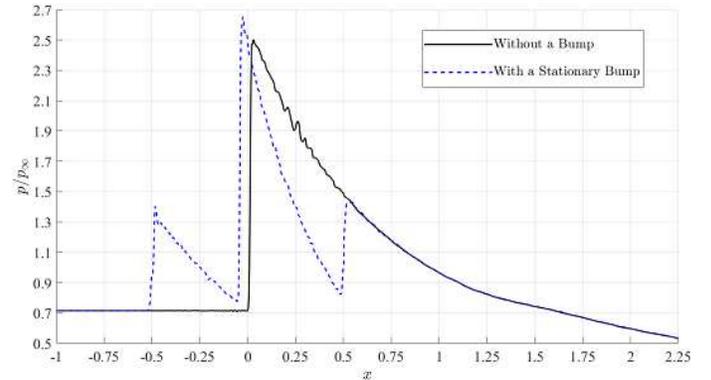}
    \caption{Comparison between the two problems, supersonic flow over a flat plate with and without a bump using the pressure distribution over the bottom boundary.}\label{fig:Fig4}
\end{figure}

Firstly, the benefit of using the bump in the SWBLI problem has to be shown before investigating the transitional effect of the morphing bump. Thus, supersonic flows over a flat plate without and with a stationary bump were simulated to obtain the steady-state solutions of the morphing bump limiting positions, as shown in Figure \ref{fig:Fig5_6}. The benefit of using the bump is indicated by plotting the pressure distribution over the lower boundary for both problems; without and with a stationary bump, as shown in Figure \ref{fig:Fig4}.

The existence of the bump created a series of weak expansion fans which in total decreased the pressure dramatically to the inlet value of pressure as shown in Figure \ref{fig:Fig4} within the region of $x\approx(0,0.5)$. Then, the pressure increased at the end of the bump to the original case (clean flat plate) at $x=0.5$, due to the generation of the trailing oblique shock. This decrease of the pressure in this simple geometry (flat plate) case shows that the bump is applicable to control the SWBLI in the airfoil case, where the pressure decreases over the upper surface, and accordingly the lift force increases. This aligns with the results of Mazaheri et al. \cite{mazaheri2015optimization}.
\begin{figure*}
    \centering
	\begin{subfigure}{\linewidth}
		\centering
		\includegraphics[width=.75\linewidth]{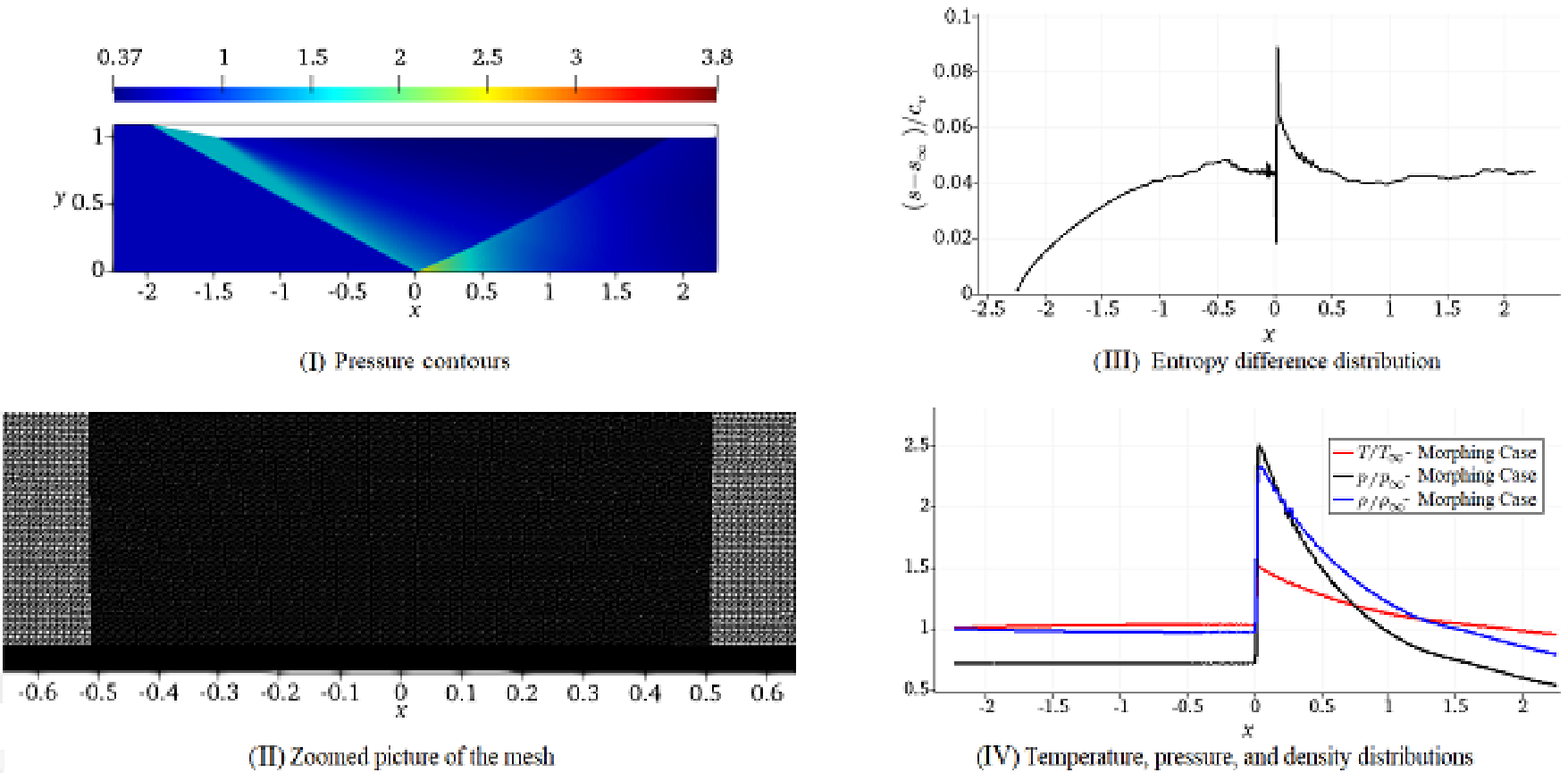}
		\caption{$t^\ast=0$ }\label{fig:Fig7}
	\end{subfigure}\\
	\begin{subfigure}{\linewidth}
		\centering
		\includegraphics[width=.75\linewidth]{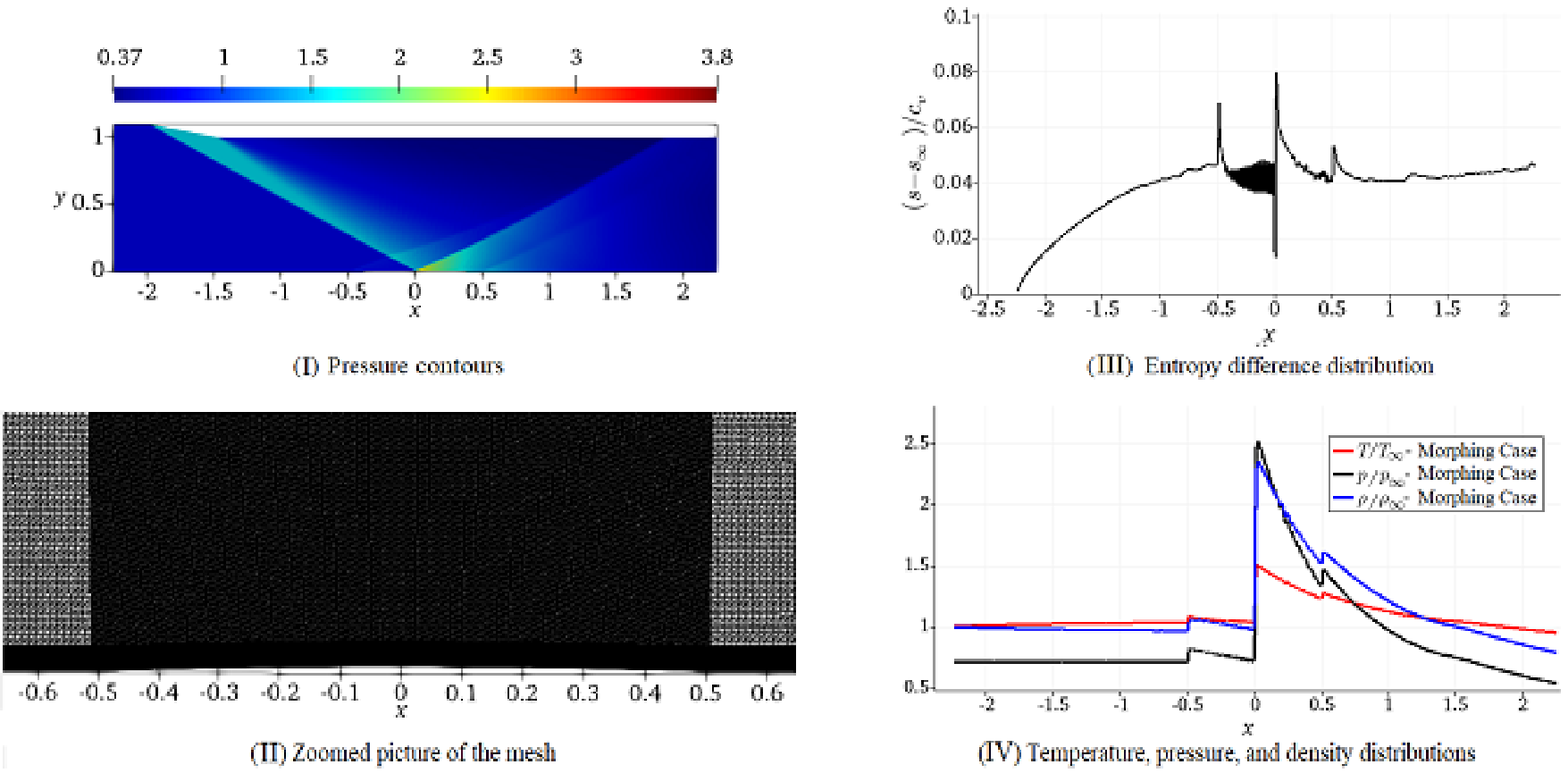}
		\caption{ $t^\ast=2$}\label{fig:Fig8}
	\end{subfigure}
	\begin{subfigure}{\linewidth}
		\centering
		\includegraphics[width=.75\linewidth]{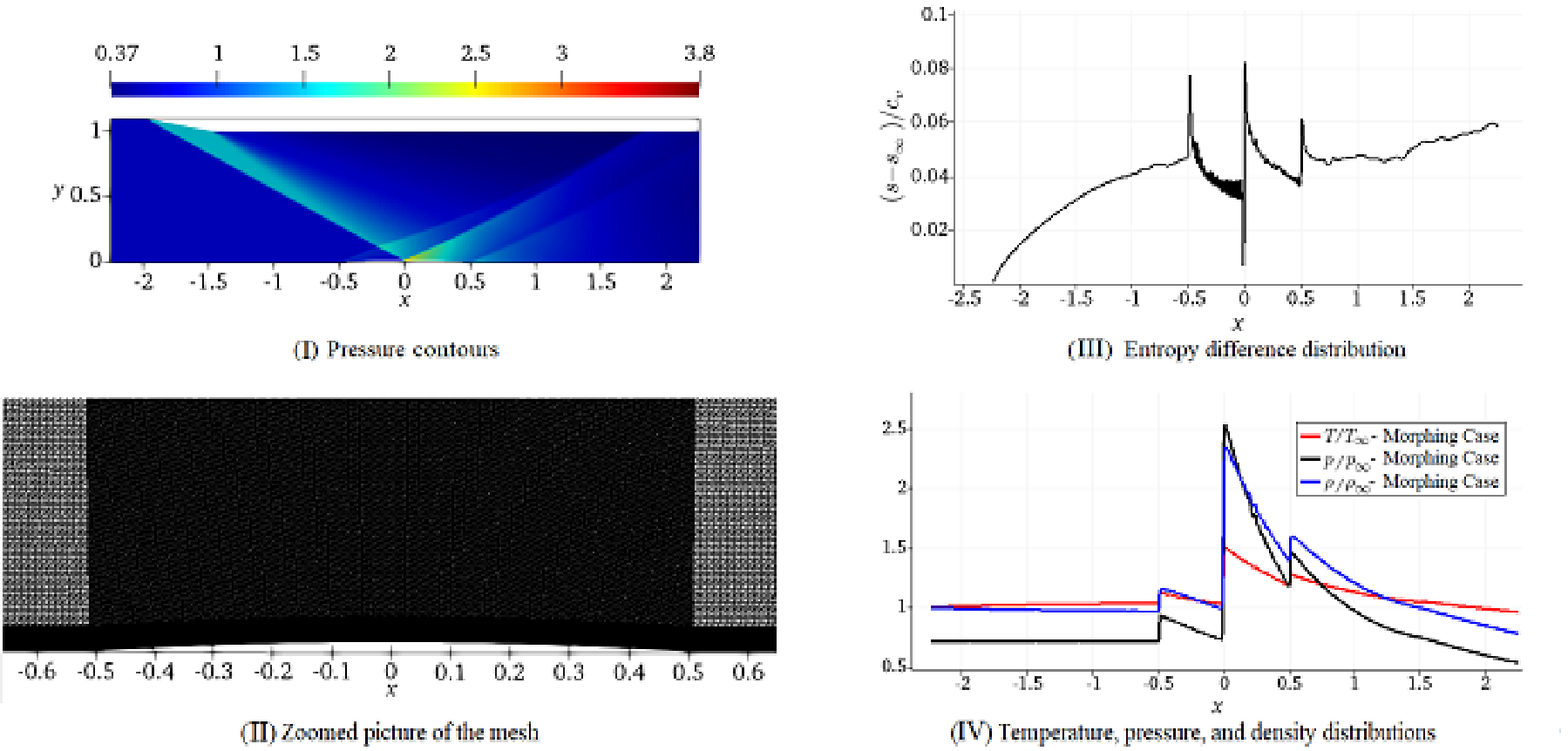}
		\caption{ $t^\ast=4$}\label{fig:Fig9}
	\end{subfigure}
\end{figure*}
\begin{figure*}
    \centering
    \ContinuedFloat
	\begin{subfigure}{.88\linewidth}
		\centering
		\includegraphics[width=.75\linewidth]{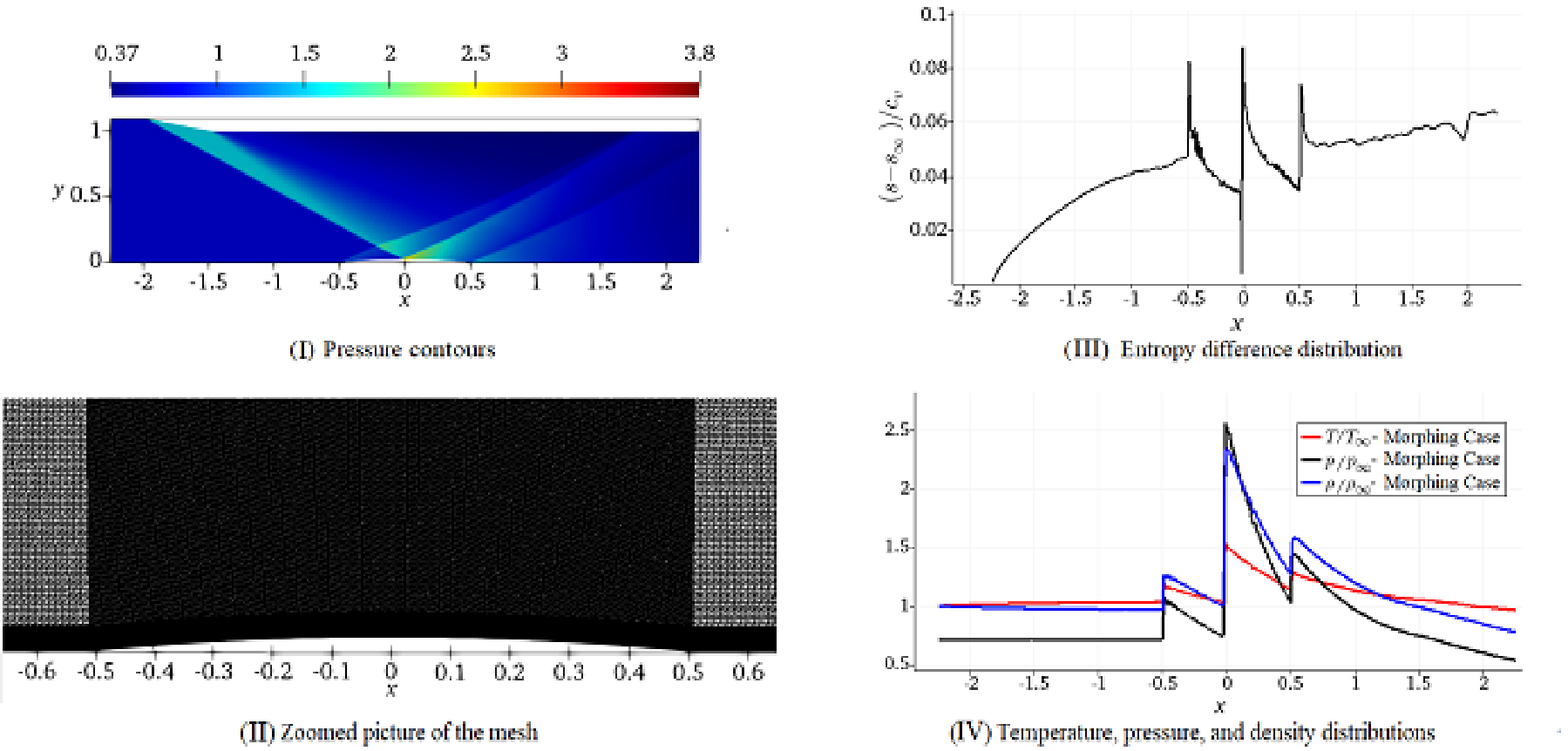}
		\caption{ $t^\ast=6$}\label{fig:Fig10}
	\end{subfigure}
	\begin{subfigure}{.88\linewidth}
		\centering
		\includegraphics[width=.75\linewidth]{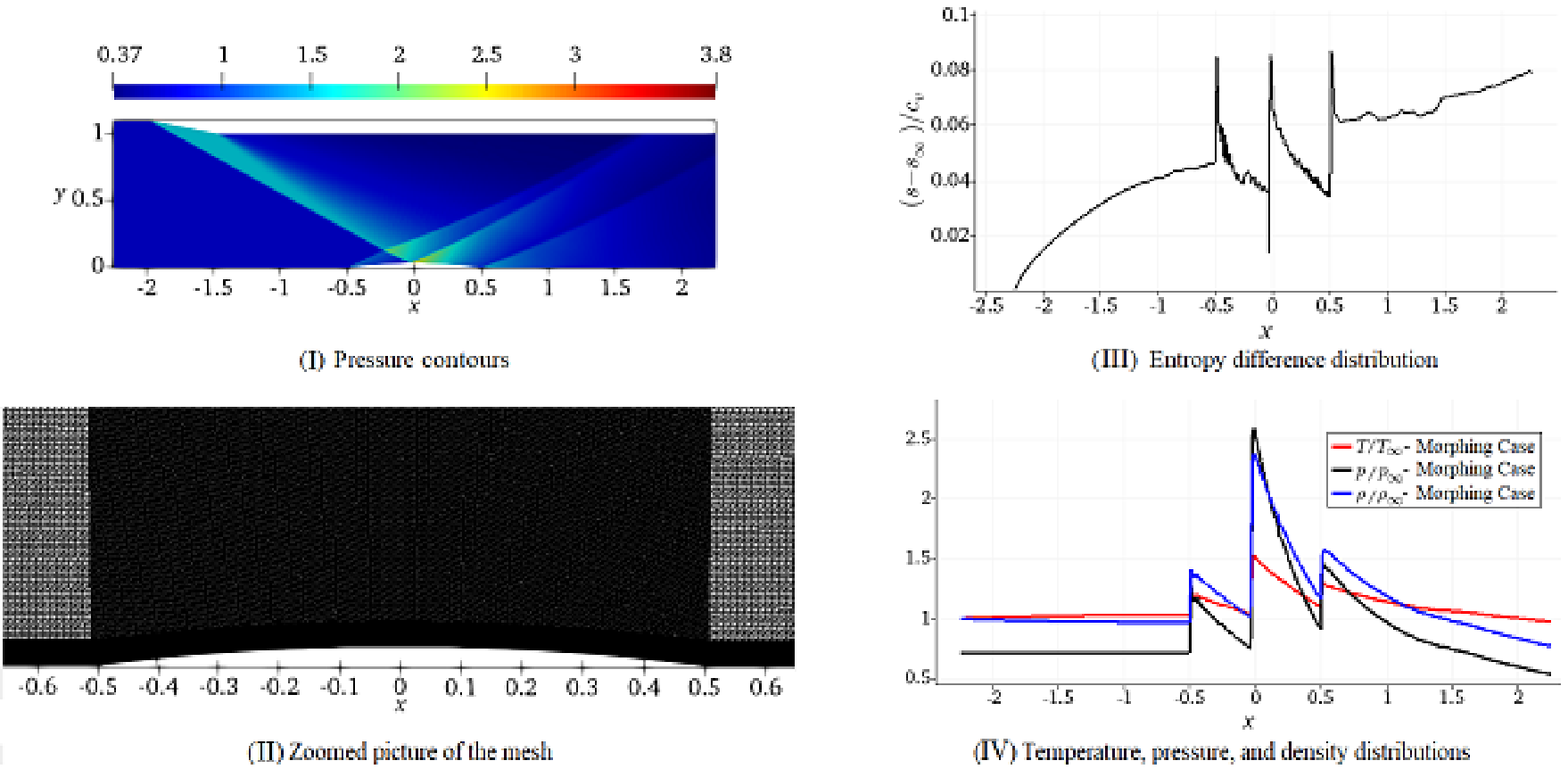}
		\caption{ $t^\ast=8$}\label{fig:Fig11}
	\end{subfigure}
	\begin{subfigure}{.88\linewidth}
		\centering
		\includegraphics[width=.75\linewidth]{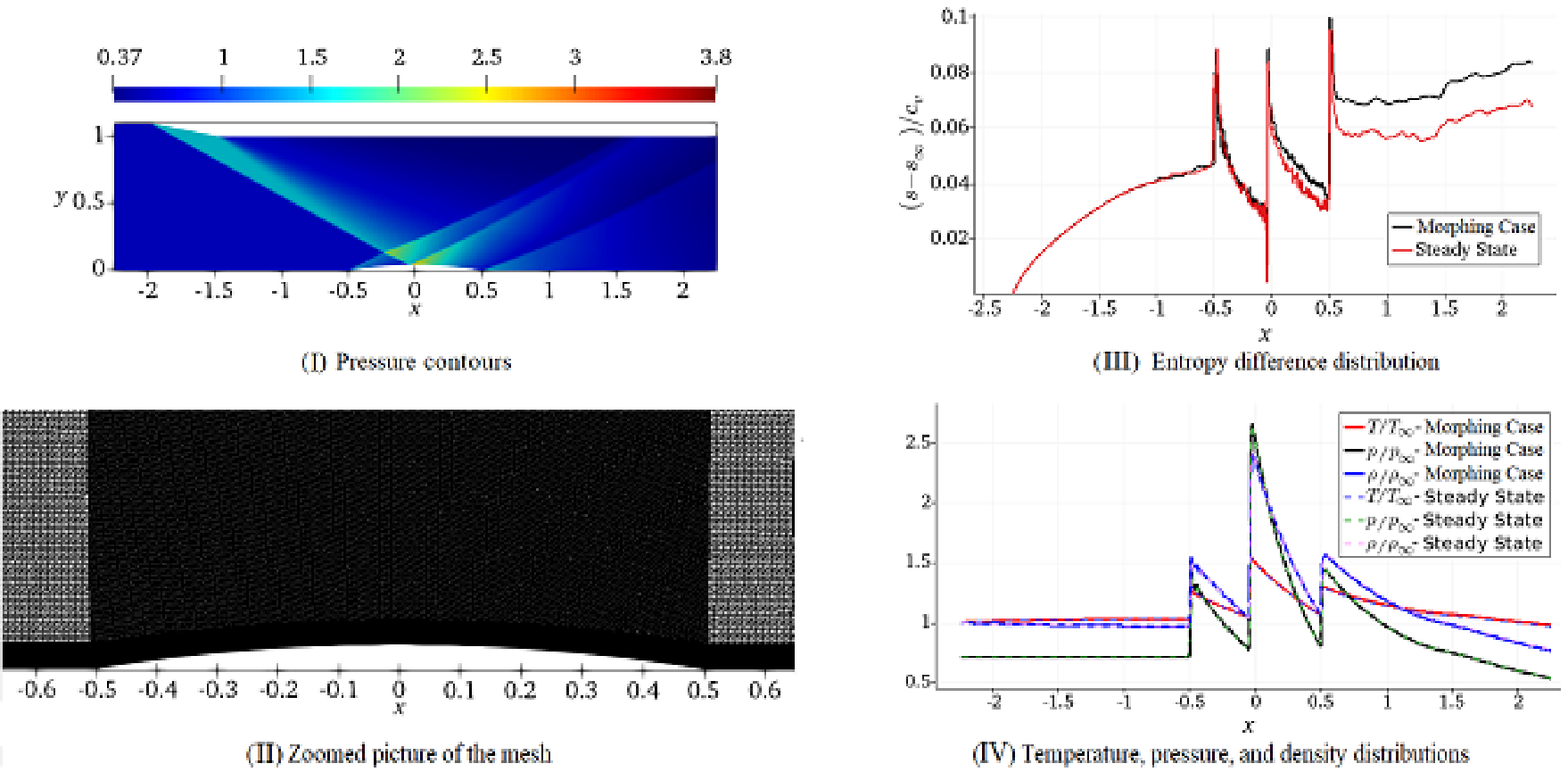}
		\caption{ $t^\ast=10$}\label{fig:Fig12}
	\end{subfigure}
	\caption{ Pressure contours, shown in (a), zoomed picture of the morphing mesh, shown in (b), and entropy, temperature, pressure, and density distributions, shown in (c) and (d), of supersonic flow over a flat plate with a morphing bump control element with constant velocity profile along the Bottom boundary.}\label{fig:Fig7_12}
\end{figure*}
\subsection{Shock Dynamics over Morphing Bump}
The transition phase of the morphing bump was investigated in detail with a constant velocity profile. The motion of the morphing bump occurred during different dimensionless times, $t^\ast_f = 10$, $7.5$, $5$, $3$, $2$, $1.8$, $1$, and $0.1$. This corresponds to a constant local Mach number, $M_b$, of $0.004$, $0.00533$, $0.008$, $0.0133$, $0.02$, $0.0222$,  $0.04$, and $0.4$ at the mid-point of the morphing bump ($x = 0$), respectively, and zero-velocity points at the two edges of the bump. Figure \ref{fig:Fig7_12} shows the transitional solution of the morphing bump with the slowest constant velocity (the Mach number at $x = 0$ is $0.004$ and the final dimensionless time, $t^\ast_f$ is $10$) by plotting the pressure contours, shown in (I), zoomed picture of the morphing mesh, shown in (II), and entropy, temperature, pressure, and density distributions, shown in (III) and (IV), along the bottom boundary at different dimensionless times, $t^\ast$. \\

\begin{figure}[H]
	\centering
	\begin{subfigure}{\linewidth}
		\centering
		\includegraphics[width=0.95\linewidth]{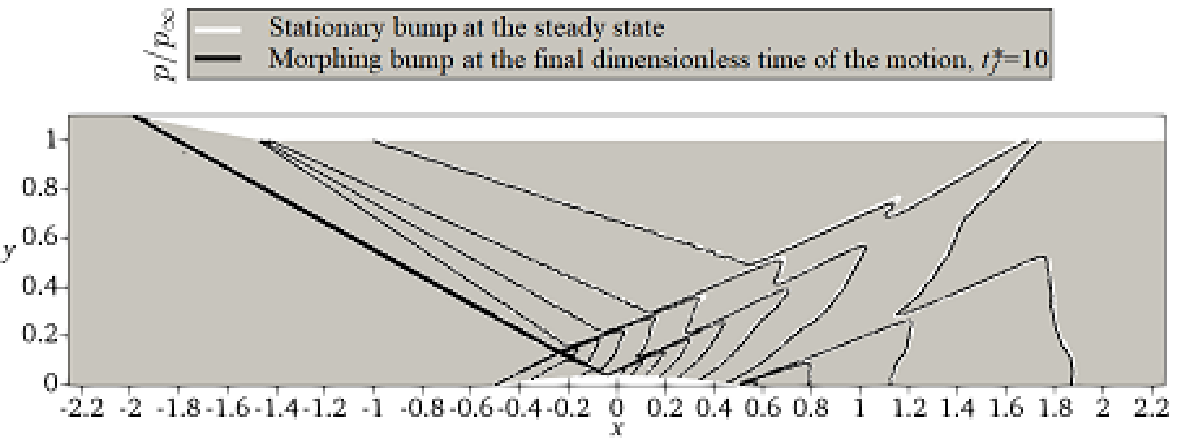}
		\caption{ Stationary bump and morphing bump at $t^{\ast}_f=10$.}\label{fig:Fig13_1}
	\end{subfigure}\\
	\begin{subfigure}{\linewidth}
		\centering
		\includegraphics[width=0.95\linewidth]{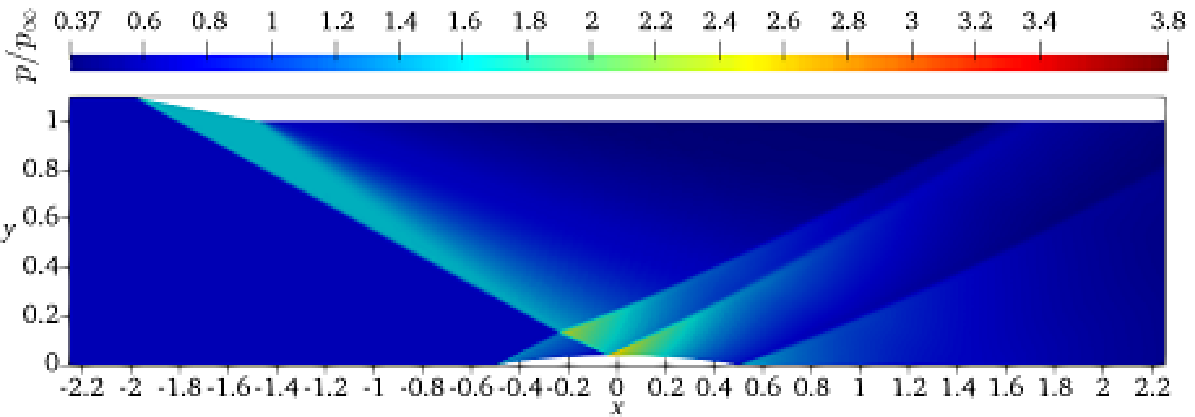}
		\caption{ Morphing bump at $t^{\ast}_f=10$.}\label{fig:Fig13_2}
	\end{subfigure}\\
	\begin{subfigure}{\linewidth}
		\centering
		\includegraphics[width=0.95\linewidth]{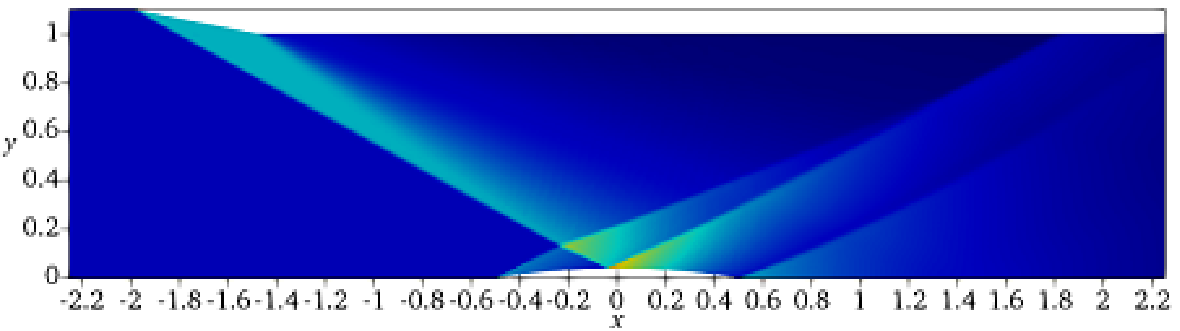}
		\caption{ Morphing bump at $t^{\ast}_f=1$.}\label{fig:Fig13_3}
	\end{subfigure}\\
	\begin{subfigure}{\linewidth}
		\centering
		\includegraphics[width=0.95\linewidth]{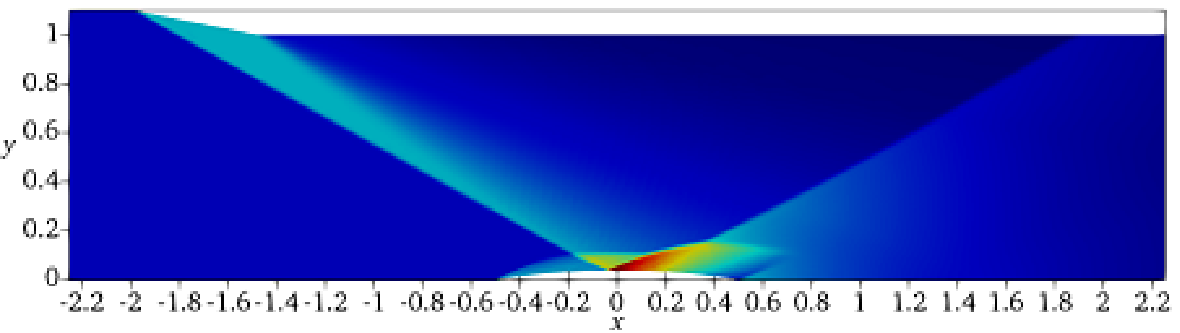}
		\caption{ Morphing bump at $t^{\ast}_f=0.1$.}\label{fig:Fig13_4}
	\end{subfigure}
	\caption{Pressure contours of supersonic flow over a flat plate with a stationary/morphing bump at different constant velocities, to show the spatial lag in the shock system.}\label{fig:Fig13}
\end{figure}

\begin{figure}[H]
	\centering
	\begin{subfigure}{\linewidth}
		\centering
		\includegraphics[width=\linewidth]{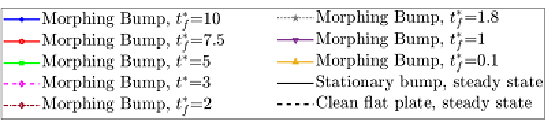}
	\end{subfigure}\\
	\begin{subfigure}{\linewidth}
		\centering
		\includegraphics[width=0.95\linewidth]{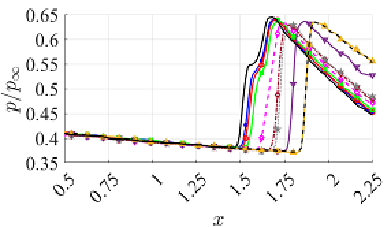}
		\caption{ Pressure distribution.}\label{fig:Fig18_1}
	\end{subfigure}\\
	\begin{subfigure}{\linewidth}
		\centering
		\includegraphics[width=0.95\linewidth]{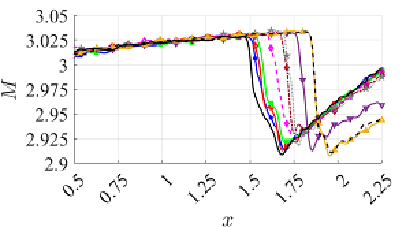}
		\caption{ Mach number distribution.}\label{fig:Fig18_2}
	\end{subfigure}
	\caption{Pressure, and Mach number distributions of supersonic flow over a flat plate with a stationary/morphing bump control element at a height of $y = 1m$, to show the change of spatial lag in the shock system at different dimensionless morphing periods.}\label{fig:Fig18}
\end{figure}

There is a spatial lag in the leading and trailing oblique shocks between the transitional solution at different values of constant velocity (different values of the final dimensionless times, $t^\ast_f$ ), and the stationary condition, as shown in Figure \ref{fig:Fig13}. This spatial lag in the shock system from the stationary condition increases when the value of the morphing bump’s constant velocity increases, i.e. the operational time of the morphing bump was reduced. This was concluded also in a rotating wedge problem, made by Margha et al. \cite{margha2021dynamic}. The spatial lag was clearly indicated by comparing the pressure, and Mach number distributions at a height of $y = 1$, between the stationary condition and the morphing conditions at different $t^{\ast}_f$, as shown in Figure \ref{fig:Fig18}. Further, the change in the spatial lag can be neglected when the operational time of the morphing bump exceeds the final dimensionless time, $t^\ast_f = 5$. 

\begin{figure}[H]
	\centering
	\begin{subfigure}{\linewidth}
		\centering
		\includegraphics[width=0.95\linewidth]{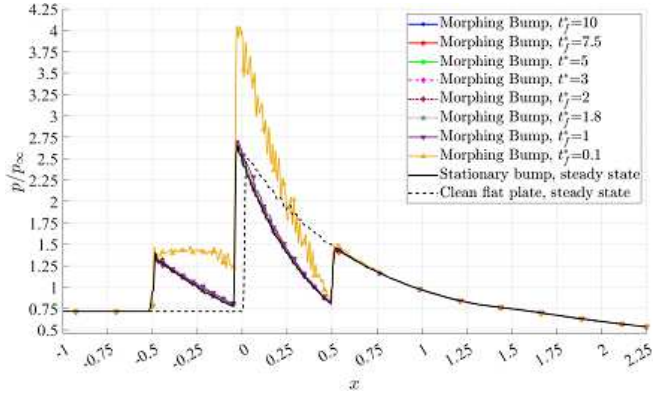}
		\caption{ Pressure distribution.}\label{fig:Fig15}
	\end{subfigure}\\
	\begin{subfigure}{\linewidth}
		\centering
		\includegraphics[width=0.95\linewidth]{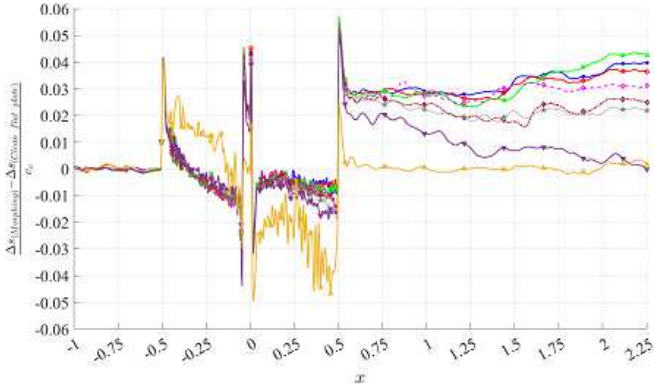}
		\caption{ Temporal change in the entropy difference distribution.}\label{fig:Fig17_1}
	\end{subfigure}
	\caption{ Pressure distribution, and the temporal change in the entropy difference distribution for the supersonic flow over a flat plate with a morphing bump over the lower boundary.}\label{fig:Fig15_17}
\end{figure}

A comparison between the start state (clean flat plate) and the final state (at the end of the morphing motion) over the bump’s surface at different motion periods was performed by obtaining the flow parameter distributions. Then, the temporal change in the entropy difference distribution with the case of a clean flat plate problem at a steady state over the lower boundary, $\left(\Delta s_{@t=t^{\ast}_f}(x) - \Delta s_{@t=0}(x)\right)/c_v$, is calculated, as shown in Figure \ref{fig:Fig15_17}. Further, the entropy difference distribution, $\Delta s(x)$, is defined as the difference between the entropy at a certain $x$ location with the inlet condition. 

When the leading oblique shock appears due to the morphing bump motion, the pressure and the density over the bump increase. Thus, the temporal change in the entropy difference rises up at the bump's leading edge. Then, the series of expansion, before and after the incident oblique shock, decrease the pressure and density along the morphing bump. This may cause negative values for the temporal change in the entropy difference over the bump at relatively low morphing speeds. Furthermore, the location of the incident shock moves slightly to the left during the upward motion of the bump. The trailing oblique shock again increases the temporal change in the entropy difference. In addition, there is a deviation between the two entropy difference over the lower boundary; the case of the stationary bump at steady state, and the case of the morphing bump at final states of the motion, see Figure \ref{fig:Fig7_12} (f) and Figure \ref{fig:Fig13} (a). The deviation happens due to the effect of spatial lag on the shock system, which results from the bump’s motion. 

\begin{figure}[H]
	\centering
	\begin{subfigure}{\linewidth}
		\centering
		\includegraphics[width=\linewidth]{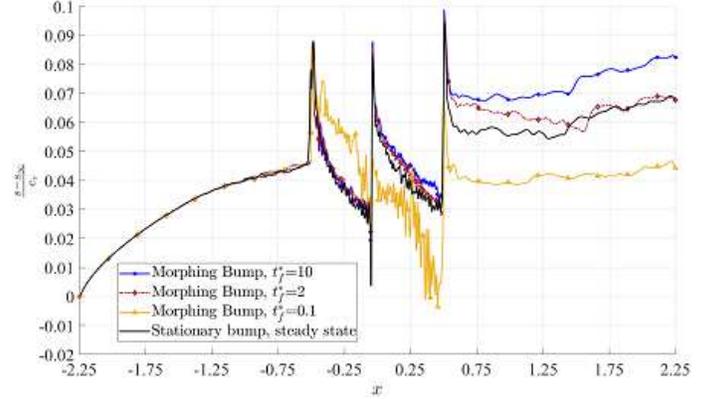}
		\caption{ Change in temporal entropy difference.}\label{fig:Fig20}
	\end{subfigure}
	\centering
	\begin{subfigure}{\linewidth}
		\centering
		\includegraphics[width=\linewidth]{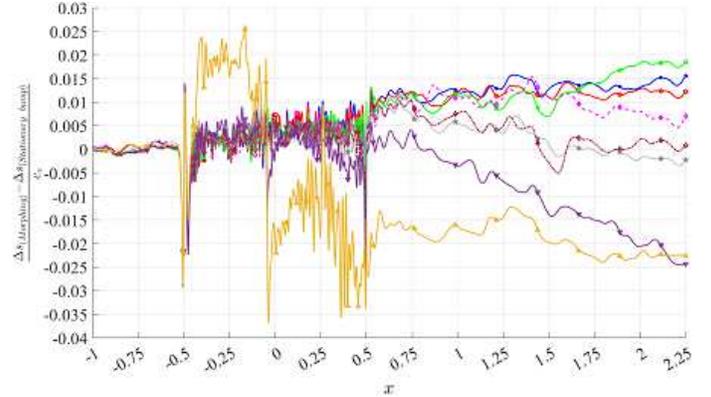}
		\caption{ Change in temporal entropy difference.}\label{fig:Fig17_2}
	\end{subfigure}
	\caption{ Pressure distribution, and the change in temporal entropy difference distribution for the supersonic flow over a flat plate with a morphing bump over the lower boundary.}\label{fig:Fig20_17}
\end{figure}

when the morphing bump is moved with a relatively high speed (small $t^{\ast}_f$), we lose the gain in the pressure difference ($p_{@t=t^{ast}_f}(x)-p_{t=0}(x)$), that is obtained from the series of expansion fans. This is explained as the pressure applied on the morphing bump would increase, which can be considered as a resistant action to the motion of the bump. Further, the compressibility effect increases with the bump’s speed, and consequently the density increases. When the air is compressed with the motion of the bump, the molecular distances decrease, and then the friction between adjacent molecules increases. Thus, the temperature increases with the bump’s motion. Thus, when the morphing bump moves faster, the flow parameters (pressure, density, and temperature) increase, as shown in Figure \ref{fig:Fig15_17} (a). This is clearly shown in the temporal change in the entropy difference distribution of Figure \ref{fig:Fig15_17} (b). Hence, the results show that the slowest velocity for the morphing bump, which approximates a quasi-steady flow during the motion, decreases the lag effect.

Despite the existence of the lag effect in the shock system, the slowest tested morphing bump's speed ($t^{\ast}_f=10$) is not the optimal one, due to the deviation in the entropy from the stationary bump at steady state, see Figure \ref{fig:Fig7_12} (f). This deviation represents the losses in the shock system, resulting from the morphing speed. Thus, comparing the entropy difference for each speed was made with the stationary bump at a steady state, as shown in Figure \ref{fig:Fig20_17}. The results show the suitable motion period, $t^{\ast}_f$, for the morphing bump at a free-stream Mach number of 2.9 is 2. The reason is that its entropy difference is very close to that of the stationary bump at a steady state. Furthermore, the lag effect in the system resulting from that morphing speed is within acceptable moderate levels, as shown in Figure \ref{fig:Fig18}.

\section{Conclusion}
The current research work aimed to investigate the morphing of a bump control element over a flat plate at a free-stream Mach number of $2.9$ and a high Reynolds number of $6.6\times10^7$. Different values of constant velocity were conducted to study the effect of the morphing bump on the shock system with non-dimensional motion periods, $t^\ast_f$ of $0.1$, $1.0$, $1.8$, $2.0$, $3.0$, $5.0$, $7.5$, $10.0$. Further, the steady state of supersonic flows over a clean flat plate and a flat plate with a stationary bump were conducted to show the beneficial effect of the bump's existence. Furthermore, a comparison between the dynamic and static cases was achieved. The results showed that a spatial lag in the shock system appears due to the dynamic motion. In addition, the lag effect remarkably increases when the local Mach number at the tip of the morphing bump, $M_b$, increases higher than $0.008$. The relatively fast morphing bump would increase the compressibility effect in the near area which may compensate for the beneficial effect of the bump’s existence. Besides, the relatively slow morphing speed results in a deviation in the entropy from the stationary bump case, which is a representation of flow momentum losses. Thus, the suitable speed to morph with is the one that results in neither a remarkable lag effect in the shock system nor high losses in the entropy deviation from the stationary steady-state case. For the case of supersonic flow with $M_\infty = 2.9$ over a flat plate, the suitable bump's morphing period, $t^{\ast}_f$, was found to be 2. For future work, various velocity profiles with different speed values are recommended to be tested for the morphing bump. This will determine the suitable morphing bump's velocity profile and speed, that achieves low time-average entropy for the flat plate problem. Then, this configuration of motion will be tested over a transonic airfoil to achieve a high time-averaged lift-to-drag ratio.
\bibliography{Hamada}

\begin{thebibliography}{27}
\providecommand{\natexlab}[1]{#1}
\providecommand{\url}[1]{\texttt{#1}}
\expandafter\ifx\csname urlstyle\endcsname\relax
  \providecommand{\doi}[1]{doi: #1}\else
  \providecommand{\doi}{doi: \begingroup \urlstyle{rm}\Url}\fi

\bibitem[Ferri and ATTI(1939)]{ferri1939experimental}
Antonio Ferri and DG~ATTI.
\newblock Experimental results with airfoils tested in the high speed tunnel at
  guidonia: Naca tm 946.
\newblock Technical report, Washington, DC: National Advisory Committee for
  Aeronautics, 1939.

\bibitem[Green(1970)]{green1970interactions}
JE~Green.
\newblock Interactions between shock waves and turbulent boundary layers.
\newblock \emph{Progress in Aerospace Sciences}, 11:\penalty0 235--340, 1970.

\bibitem[Dolling(2001)]{dolling2001fifty}
David~S Dolling.
\newblock Fifty years of shock-wave/boundary-layer interaction research: what
  next?
\newblock \emph{AIAA journal}, 39\penalty0 (8):\penalty0 1517--1531, 2001.

\bibitem[Bushnell(2004)]{bushnell2004shock}
Dennis~M Bushnell.
\newblock Shock wave drag reduction.
\newblock \emph{Annu. Rev. Fluid Mech.}, 36:\penalty0 81--96, 2004.

\bibitem[Ashill and Fulker(1992)]{ashill199292}
PR~Ashill and JL~Fulker.
\newblock 92-01-022 a novel technique for controlling shock strength of
  laminar-flow aerofoil sections.
\newblock \emph{DGLR BERICHT}, pages 175--175, 1992.

\bibitem[Milholen and Owens(2005)]{milholen2005application}
William Milholen and Lewis Owens.
\newblock On the application of contour bumps for transonic drag reduction.
\newblock In \emph{43rd AIAA Aerospace Sciences Meeting and Exhibit}, page 462,
  2005.

\bibitem[P{\"a}tzold et~al.(2006)P{\"a}tzold, Lutz, Kramer, and
  Wagner]{patzold2006numerical}
Martin P{\"a}tzold, Thorsten Lutz, Ewald Kramer, and Siegfried Wagner.
\newblock Numerical optimization of finite shock control bumps.
\newblock In \emph{44th AIAA Aerospace Sciences Meeting and Exhibit}, page
  1054, 2006.

\bibitem[Fulker(1999)]{fulker1999euroshock}
J~Fulker.
\newblock The euroshock programme (a european programme on active and passive
  control of shock waves).
\newblock In \emph{17th Applied Aerodynamics Conference}, page 3174, 1999.

\bibitem[Sommerer et~al.(2000)Sommerer, Lutz, and
  Wagner]{sommerer2000numerical}
Andreas Sommerer, Thorsten Lutz, and Siegfried Wagner.
\newblock Numerical optimisation of adaptive transonic airfoils with variable
  camber.
\newblock In \emph{Proceedings of the 22nd International Congress of the
  Aeronautical Sciences}, volume~27. Optimage Ltd. Edinburgh, UK, 2000.

\bibitem[Wong et~al.(2007)Wong, Le~Moigne, and Qin]{wong2007parallel}
WS~Wong, A~Le~Moigne, and N~Qin.
\newblock Parallel adjoint-based optimisation of a blended wing body aircraft
  with shock control bumps.
\newblock \emph{The Aeronautical Journal}, 111\penalty0 (1117):\penalty0
  165--174, 2007.

\bibitem[Tian et~al.(2014)Tian, Liu, and Li]{tian2014multi}
Yun Tian, PeiQing Liu, and Zhi Li.
\newblock Multi-objective optimization of shock control bump on a supercritical
  wing.
\newblock \emph{Science China Technological Sciences}, 57\penalty0
  (1):\penalty0 192--202, 2014.

\bibitem[Mazaheri et~al.(2015)Mazaheri, Kiani, Nejati, Zeinalpour, and
  Taheri]{mazaheri2015optimization}
K~Mazaheri, KC~Kiani, A~Nejati, M~Zeinalpour, and R~Taheri.
\newblock Optimization and analysis of shock wave/boundary layer interaction
  for drag reduction by shock control bump.
\newblock \emph{Aerospace Science and Technology}, 42:\penalty0 196--208, 2015.

\bibitem[Eastwood and Jarrett(2012)]{eastwood2012toward}
Jeremy~P Eastwood and Jerome~P Jarrett.
\newblock Toward designing with three-dimensional bumps for lift/drag
  improvement and buffet alleviation.
\newblock \emph{AIAA journal}, 50\penalty0 (12):\penalty0 2882--2898, 2012.

\bibitem[Tian et~al.(2011)Tian, Liu, and Feng]{tian2011shock}
Yun Tian, PeiQing Liu, and PeiHua Feng.
\newblock Shock control bump parametric research on supercritical airfoil.
\newblock \emph{Science China Technological Sciences}, 54\penalty0
  (11):\penalty0 2935, 2011.

\bibitem[Monner et~al.(2000)Monner, Breitbach, Bein, and
  Hanselka]{monner2000design}
HP~Monner, E~Breitbach, Th~Bein, and H~Hanselka.
\newblock Design aspects of the adaptive wing—the elastic trailing edge and
  the local spoiler bump.
\newblock \emph{The Aeronautical Journal}, 104\penalty0 (1032):\penalty0
  89--95, 2000.

\bibitem[Olivett et~al.(2021)Olivett, Corrao, and Karami]{olivett2021flow}
Anthony Olivett, Peter Corrao, and M~Amin Karami.
\newblock Flow control and separation delay in morphing wing aircraft using
  traveling wave actuation.
\newblock \emph{Smart Materials and Structures}, 30\penalty0 (2):\penalty0
  025028, 2021.

\bibitem[Elbadry et~al.(2021)Elbadry, Guaily, Boraey, and
  AbdelRahman]{elbadry2021active}
Yusuf~T Elbadry, Amr~Gamal Guaily, Mohammed~A Boraey, and Mohamed~M
  AbdelRahman.
\newblock Active morphing control of airfoil at low reynolds number using
  level-set method.
\newblock In \emph{2021 3rd Novel Intelligent and Leading Emerging Sciences
  Conference (NILES)}, pages 265--268. IEEE, 2021.

\bibitem[Botez et~al.(2018)Botez, Koreanschi, Gabor, Tondji, Guezguez,
  Kammegne, Grigorie, Sandu, Mebarki, Mamou, et~al.]{botez2018numerical}
RM~Botez, A~Koreanschi, OS~Gabor, Y~Tondji, M~Guezguez, JT~Kammegne,
  LT~Grigorie, D~Sandu, Y~Mebarki, M~Mamou, et~al.
\newblock Numerical and experimental transition results evaluation for a
  morphing wing and aileron system.
\newblock \emph{The Aeronautical Journal}, 122\penalty0 (1251):\penalty0
  747--784, 2018.

\bibitem[Popov et~al.(2010)Popov, Grigorie, Botez, Mamou, and
  M{\'e}barki]{popov2010real}
Andrei~V Popov, Lucian~T Grigorie, Ruxandra Botez, Mahmood Mamou, and Youssef
  M{\'e}barki.
\newblock Real time morphing wing optimization validation using wind-tunnel
  tests.
\newblock \emph{Journal of Aircraft}, 47\penalty0 (4):\penalty0 1346--1355,
  2010.

\bibitem[Bruce and Colliss(2015)]{bruce2015review}
PJK Bruce and SP~Colliss.
\newblock Review of research into shock control bumps.
\newblock \emph{Shock Waves}, 25\penalty0 (5):\penalty0 451--471, 2015.

\bibitem[Ma et~al.(2014)Ma, Lu, Fang, and Wang]{ma2014study}
Li~Ma, Lipeng Lu, Jian Fang, and Qiuhui Wang.
\newblock A study on turbulence transportation and modification of
  spalart--allmaras model for shock-wave/turbulent boundary layer interaction
  flow.
\newblock \emph{Chinese Journal of Aeronautics}, 27\penalty0 (2):\penalty0
  200--209, 2014.

\bibitem[Kurganov and Tadmor(2000)]{kurganov2000new}
Alexander Kurganov and Eitan Tadmor.
\newblock New high-resolution semi-discrete central schemes for
  hamilton--jacobi equations.
\newblock \emph{Journal of Computational Physics}, 160\penalty0 (2):\penalty0
  720--742, 2000.

\bibitem[Kurganov et~al.(2001)Kurganov, Noelle, and
  Petrova]{kurganov2001semidiscrete}
Alexander Kurganov, Sebastian Noelle, and Guergana Petrova.
\newblock Semidiscrete central-upwind schemes for hyperbolic conservation laws
  and hamilton--jacobi equations.
\newblock \emph{SIAM Journal on Scientific Computing}, 23\penalty0
  (3):\penalty0 707--740, 2001.

\bibitem[Greenshields et~al.(2010)Greenshields, Weller, Gasparini, and
  Reese]{greenshields2010implementation}
Christopher~J Greenshields, Henry~G Weller, Luca Gasparini, and Jason~M Reese.
\newblock Implementation of semi-discrete, non-staggered central schemes in a
  colocated, polyhedral, finite volume framework, for high-speed viscous flows.
\newblock \emph{International Journal for Numerical Methods in Fluids},
  63\penalty0 (1):\penalty0 1--21, 2010.

\bibitem[Marcantoni et~al.(2012)Marcantoni, Tamagno, and
  Elaskar]{marcantoni2012high}
Luis F~Guti{\'e}rrez Marcantoni, Jos{\'e}~P Tamagno, and Sergio~A Elaskar.
\newblock High speed flow simulation using openfoam.
\newblock \emph{Mec{\'a}nica Computacional}, 31\penalty0 (16):\penalty0
  2939--2959, 2012.

\bibitem[Hirt et~al.(1974)Hirt, Amsden, and Cook]{hirt1974arbitrary}
Cyrill~W Hirt, Anthony~A Amsden, and JL~Cook.
\newblock An arbitrary lagrangian-eulerian computing method for all flow
  speeds.
\newblock \emph{Journal of computational physics}, 14\penalty0 (3):\penalty0
  227--253, 1974.

\bibitem[Margha et~al.(2021)Margha, Hamada, Knight, and
  Eltaweel]{margha2021dynamic}
Lubna Margha, Ahmed~A Hamada, Doyle~D Knight, and Ahmed Eltaweel.
\newblock Dynamic transition from regular to mach reflection over a moving
  wedge.
\newblock In \emph{ASME International Mechanical Engineering Congress and
  Exposition}, volume 85581, page V004T04A006. American Society of Mechanical
  Engineers, 2021.

\end{thebibliography}
\end{multicols*}
\end{document}